# Critical Current Measurement of REBCO Cables by Using a Superconducting Transformer

H. Yu, J. Lu, J. D. Weiss, and D. C. van der Laan

*Abstract*—Development of REBCO cables that carry high electrical current in high magnetic field is crucial for future large-scale magnet applications. This experimental work presents the critical current measurements of two different REBCO cables by a test facility at the National High Magnetic Field Laboratory (NHMFL). The simple-stacked cable is made by the NHMFL by stacking 21 REBCO tapes without soldering. The Conductor-on-Round-Core (CORC®) cable provided by Advanced Conductor Technologies has 21 layers of REBCO tapes with 2 tapes/layer. The test facility consists of a 12 T split solenoid magnet with 15 cm bore providing transverse field to the samples, a superconducting transformer (SCT) as a current source providing up to 45 kA current. Special attentions were paid to fabrication of solder joints between REBCO cables and the SCT output. The voltage-current traces were measured as a function of magnetic field at 4.2 K, from which the critical currents are determined. The details of this measurement are discussed.

## I. Introduction

For large superconducting magnet systems, such as nuclear fusion and particle accelerator magnets, development of superconducting cables that carry tens of kA of current is essential. Cable made with high temperature superconductor (HTS), in particular, is a key component of future ultrahigh field magnet. REBCO coated conductor, a HTS tape that allows very high current density at liquid helium temperatures and very high fields, is an excellent candidate for making large HTS cables. In the past decade, a few types of REBCO cables have been developed [1]-[4]. There are strong interests in continuing the development of these REBCO cables for both nuclear fusion and particle accelerator magnet applications [5], [6]. In addition, a new concept of the so-called integrated-coil-form magnet was recently proposed based on a conductor with multiple simply stacked REBCO tapes [7].

For the development of HTS cables, a facility for measuring high critical current ($I_c$) in high magnetic field is critically important. The National High Magnetic Field Laboratory (NHMFL) has a facility with a 12 T split solenoid magnet and a recently developed superconducting transformer (SCT) that provide current up to 45 kA [8], [9].

In this paper, critical currents of a simple-stacked REBCO cable and a Conductor-on-Round-Core (CORC®) cable were measured at 4.2 K in magnetic field up to 12 T. The experimental details and results will be presented and discussed.

## II. Experiment

### A. The Simple-stacked Cable Sample

The REBCO tapes in this cable is 4 mm wide SuperPower SCS4050-AP. They are from 5 different production batches with $I_c$ of 116 - 128 A at 77 K in self-field. The simple-stacked cable was made at the NHMFL. A 4.0 mm wide and 4.3 mm deep U-shaped groove was machined in the G-10 base of the test probe as depicted in Fig. 1. A total of 21 tapes were placed. The remaining space of the groove was then filled with lint-free papers which provide a transverse pressure to ensure good electrical contact between tapes. During the $I_c$ measurement, the ab-plane of REBCO tapes was parallel to the applied magnetic field. The polarity of the current was such that the electromagnetic force pushed the tapes to the center of the probe. This force also provided additional contact pressure between tapes. For instance, at 11.5 T and 10 kA, the electromagnetic force results in 1.4 to 28.8 MPa contact pressure from the out most to the inner most tape. This is sufficient to ensure good electrical contact [10] between tapes to allow reasonable current redistribution in the cable. Voltage taps V1 – V4 were soldered using $Pb_{37}Sn_{63}$ solder as shown in Fig. 1.

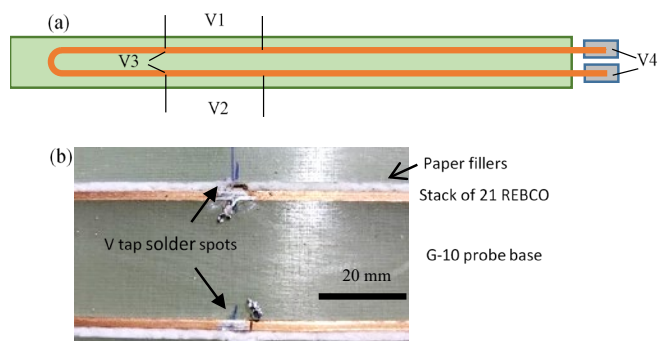

Fig. 1. (a) a schematic of the simple-stacked REBCO cable sample (not to scale) and voltage tap locations. (b) a picture of the post-test simple-stacked cable. REBCO tapes sits in the grooves on the probe. Both voltage taps came off during testing.

This work is supported by the user collaboration grant program (UCGP) of the NHMFL which is supported by NSF through NSF-DMR-1157490 and 1644779, and the State of Florida; the Department of Energy through award DE-SC0013723, DE-SC14009 and DE-SC0018127. *(Corresponding author: Jun Lu.)*
H. Yu and J. Lu are with the National High Magnetic Laboratory, Tallahassee, FL 32310, USA (e-mail: junlu@magnet.fsu.edu).

The joints to SCT were 50 mm long with edges of REBCO tapes being soldered to the broad face of the SCT output (NbTi Rutherford cables). To increase the joint area, an oxy-gen-free copper strip of $4 \times 50 \times 0.25$ mm$^3$ was inserted be-tween each pair of adjacent tapes and soldered together, as shown in Fig. 2. An aluminum heater block under the SCT output cable and a thermocouple were used to control and monitor the soldering temperature. The solder material was $Pb_{37}Sn_{63}$. The soldering procedure for each joint was at about 210 °C, and lasted less than 15 min. Under this heating condition, the REBCO $I_c$ degradation and the associated joint resistance rise should be small [11]. In addition, the joints were in low field region where the critical current is high, small degradation would not likely compromise the experiment.

### B. The CORC® Cable Sample

The CORC® cable consists of 42 SuperPower REBCO tapes. It was manufactured at the Advanced Conductor Technologies (ACT), where the test sample was also prepared and installed on the test probe. The geometric dimensions of the CORC® ca-ble are summarized in Table I. Two sections of the CORC® ca-ble were soldered to a 20 cm long copper block which would be outside the high field region for applied cur-rent return. Each of the other ends of the two sections was soldered to a 20 cm long copper channel where a NbTi Rutherford cable was also sol-dered as a jumper for connecting to the SCT. Pure indium (melt-ing point 156 °C) was used as sol-der to minimize the possible thermal degradation of the REBCO tapes and to minimize the joint resistance [11]. The main body of the probe was made of a G-10 base with machined grooves where the CORC® cables sit and a G-10 cover was bolted to the base for mechanical sta-bility during critical current measurements. A picture of the sample probe for the CORC® sample is shown in Fig. 3(a). A few voltage taps were soldered as shown in Fig. 3(b). V1 and V2 are for leg A; V3 and V4 are for leg B. Since these voltage taps are located at the transition area between the copper and the CORC® cable, it is possible to capture voltages due to the current transition from copper to the superconducting cable. It will not, however, interfere with the determination of the criti-cal current. V5 is for measuring the resistance in the solder joints on the copper return block, and V6 is for the entire sample including the solder joint between the SCT and the probe, which is made with $Pb_{37}Sn_{63}$ solder. In addition, a calibrated Hall sen-sor was placed between leg A and B to measure the actual mag-netic field including contribution from self-field.

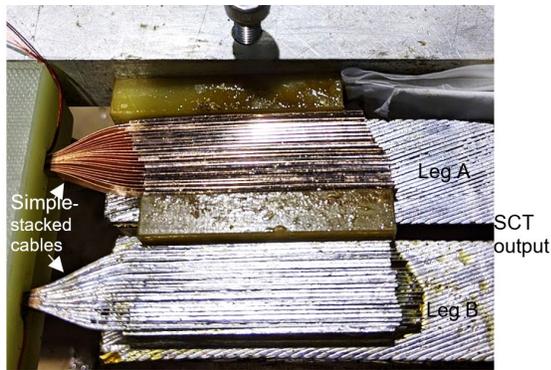

Fig. 2. Fabrication of low resistance joint between the simple-stacked cable and SCT output. Before soldering (Leg A) and after solder (Leg B).

TABLE I
CORC® CABLE PARAMETERS

| Specifications | Values |
| --- | --- |
| # of tapes | 42 |
| # of layers | 21 |
| Substrate thickness (μm) | 30 |
| Copper stabilizer thickness (μm) | 5 |
| Tape width (mm) | 3, 4 |
| Polyester insulation thickness (μm) | 30 |
| Former diameter (mm) | 2.75 |
| Cable outer diameter (mm) | 4.73 |
| $I_c$ at 77 K self-field (A) | 4809 |
| Cable Je at 77 K self-field (A/mm2) | 263 |

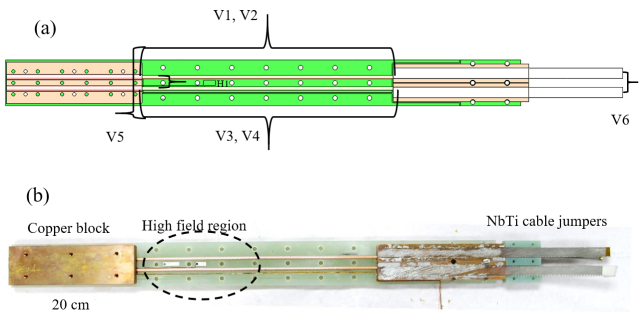

Fig. 3. (a) A schematic indicating the locations of voltage taps. A Hall sensor H1 is placed between the two legs to measure the actual magnetic field. (b) A picture of the CORC® cable sample. 20 cm long copper terminals were soldered by pure indium.

### C. Critical Current Measurement

The magnetic field was provided by an Oxford 12 T split magnet (Fig. 4). It has a 15 cm horizontal bore. The test probes with a cross-section of $30 \times 70$ mm$^2$ were inserted into the mag-net vertically. The measurements were performed in liquid he-lium.

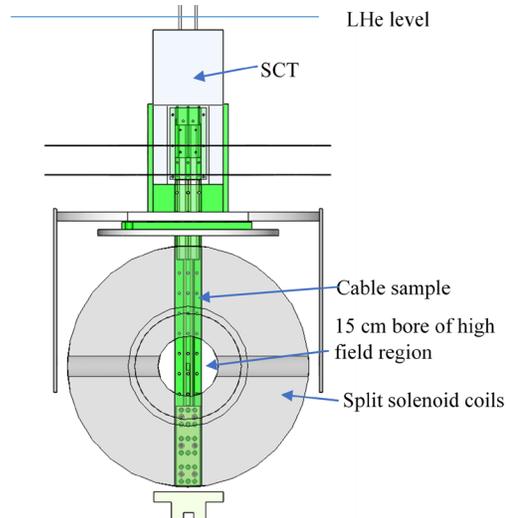

Fig. 4. Schematic of the superconducting cable sample in the NHMFL 12 T split magnet

The SCT was soldered to the sample. Then the SCT and the sample assembly was inserted into the magnet. A Rogowski coil as a part of the SCT was used to measure the output current by integrating its inductive voltage. Additional output current monitoring was accomplished by two Hall sensors each posi-tioned at the center of an output cable [9].

A Cryomagnetics model 4G magnet power supply was used to drive the primary current of SCT. This power supply also provided automatic quench detection and protection of the SCT or the sample. Data acquisition was done by a National Instru-ments SCXI-1000 with two NI SCXI-1125 input mod-ules and a LabVIEW program.

For the simple-stacked cable sample, the applied field and current polarity was such that electromagnetic force pulled the two legs towards each other. For the CORC® cable sample, however, the polarity was such that electromagnetic force pushes the two legs apart. The advantage of this configuration was that the self-field was added to the applied field.

The experiment began with ramping the SCT primary current to 20 A (corresponding SCT output of about 14 kA) in zero field and held for a few minutes for joint resistance measurements. Subsequently, the magnet was ramped to different fields for $I_c$ measurements. Before each measurement, the current induced by the field ramp was quenched by energizing quench heaters located on the secondary coil of the SCT. Then the current was ramped at about 300 A/s (SCT primary at 0.4 A/s) while various voltages were recorded until the electrical field was above 1 µV/cm or a quench was detected. The critical currents and n values were determined by fitting the $E - I$ curves with

$$E = E_c (I / I_c)^n \tag{1}$$

where $E_c$ is the criterion of 1 µV/cm.

## III. RESULTS AND DISCUSSIONS

The CORC® sample and the simple-stacked sample were tested in a single magnet cooldown campaign. The CORC® sample was cooled down with the magnet and tested. After the test, the CORC® sample and the SCT assembly was removed from the magnet cryostat. Subsequently the simple-stacked sample was soldered to the SCT and inserted into the cold mag-net for testing.

### A. The Simple-stacked cable

The joint resistance between the simple-stacked sample and the SCT is obtained from slopes of the V4-I traces before the resistive transitions to be 7.4 ± 0.4 nΩ. Independently, the sam-ple current decay time constant $\tau$ was measured to determine the joint resistance $R_j$ by with

$$R_j = L / \tau \tag{2}$$

where $L$ is the inductance of the circuit, which is the sum of SCT secondary inductance of 2.53 µH [8] and the sample in-ductance of 0.98 µH which is obtained from the measured in-ductive voltages during current ramps. With the measure $\tau$ = 520 s, the joint resistance is calculated to be 6.8 ± 0.4 nΩ, in reasonable agreement with that from the V4-I slopes. The re-sistance of the joint can be analyzed by a model of interfacial resistors and copper resistors network. From this analysis we obtain a REBCO interfacial resistivity of about 120 nΩ-cm², significantly higher than the optimum value of 25 nΩ-cm² re-ported in similar REBCO tapes [11]. This is indicative of ex-cessive heating during the joint fabrication process. Signifi-cantly lower joint resistance may be achieved by minimizing joint soldering temperature and duration.

$E-I$ traces obtained from V1 tap at 5 T, 6 T and 7 T are shown in Fig.5 where the background is subtracted for each trace. The fits by equation (1) are also plotted as solid lines. The critical currents are determined to be 16.2 kA ($n$ = 12.5), 17.1 kA ($n$ = 16), and 17.2 kA ($n$ = 13) for 5 T, 6 T and 7 T respectively.

Above 7 T, both V1 and V2 taps failed. So V4-I traces, which have significant resistive and inductive backgrounds, are ana-lyzed to determine critical currents. The electrical field E4 is defined as V4 divided by the total length of high field region of both legs of 30 cm. It is noted that all E4-I curves (not shown) have $n$ values of about 5, considerably lower than those from V1-I traces. Consequently, the critical currents obtained by V4-I are lower. For instance, the critical current determined by V4-I at 6 T is 16.7 kA, 0.4 kA lower than that determined by V1-I. Nevertheless, critical currents by V4-I correspond well with the true $I_c$ and are valuable for characterization of the sample. To distinguish critical current by V4-I from the true $I_c$, we denote $I_c^*$ as the critical current by V4-I. $I_c^*$ versus applied field is plotted in Fig. 6. As expected for REBCO with its ab-plane parallel to magnetic field, $I_c^*$ is not a strong function of field. In decreasing field from 12 T down, $I_c^*$ is almost independent of field at about 12 kA which is equivalent of about 580 A for each tape. The significantly lower $I_c^*$ in decreasing field compared with that in increasing field suggest some irreversible degradation possibly by electromagnetic stress during the preceding measurements.

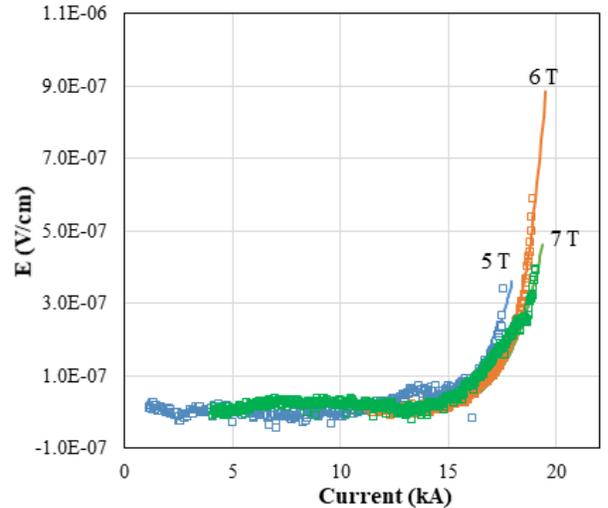

Fig. 5.  $E-I$ curves of the simple stacked sample at applied field of 5 T, 6 T and 7 T. The solid lines are the power law fits.

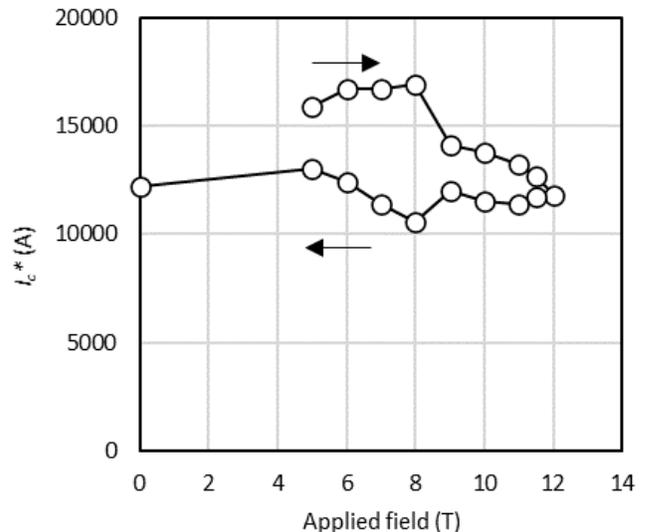

Fig. 6.  $I_c^*$ of the simple stacked sample as a function of applied magnetic field. The solid line is a guide to the eye.

### B. The CORC® Sample

The total joint resistance is 24.1 nΩ obtained by fitting the V-I data of V6 tap with a straight line. This includes contributions from joints connecting SCT and the sample, the solder joints at both ends of each leg, and the copper bottom terminal. This value agrees with what is obtained by the SCT output current decay time measurement.

The resistance of the return joints measured by V5 is 15.6 nΩ. This relatively high joint resistance is likely due to two mechanisms. First, the Cu used for the joint was found have a resistivity of 1.69 nΩ-m at 4.2 K, which is about 8 times higher than it should be for non-annealed OFHC copper. Second, complications during manufacturing required a heating time of over 15 minutes, which is enough time to significantly impact the contact resistance between soldered REBCO connections [11].

It should be noted that during the current ramp, V1, V3, V4 increased with current in a rate equivalent to a resistance of 1.6, 3.3 and 3.1 nΩ respectively. This resistance comes from the location of the voltage taps soldered within the joint structures, and was not observed in NbTi cable nor in the simple-stacked cable where voltage taps were located directly on the superconducting cables. In addition, the return resistance increased from 15.6 to 25.1 nΩ when the magnetic field was increased from 0 to 11.5 T which is likely due to the magnetoresistance of copper and solder.

The E-I curves from 5 to 11.5 T are shown in Fig. 7. The E-I curve at applied field of 5 T is fitted by equation (1) with $I_c$ = 23.7 kA, $n$ = 9. In subsequent tests at higher fields, the transitions have much higher $n$ values than those in previous CORC® cable measurements [12], [13]. In addition, $n$ value increases with applied magnetic field. This is different from those of a single tape where $n$ value decreases with magnetic field. So the observed sharp transitions might be thermal run-aways (quenches) that caused by damages occurred after the test at 5 T.

Interestingly for the E-I curves with quenches, it took up to 2 seconds (0.15 s per data point) for the electric field to develop above the criterion. It is much slower than that of low $T_c$ superconducting cables. This seems to be consistent with relatively low quench propagation velocity of REBCO tapes [14]. Meanwhile, the current redistribution between REBCO tapes does not seem to be sufficient to prevent the quench. In other words, once some damaged tapes go through early transitions, the current cannot sufficiently transfer to other tapes to prevent hotspots on the damaged tapes. Current redistribution is through inter-tape contacts along the cable and at the soldered terminals. In our case, since the inter-tape contact resistance along the cable is likely to be much higher than that at the terminal solder joints [15],[16], we can assume the contact current redistribution mostly occurred at the terminals. The high terminal joint resistance as measured in this sample hindered sufficient current sharing between REBCO tapes, failed to prevent the thermal run-aways.

The quenches after the 5 T test are likely due to the mechanical damage by transverse electromagnetic force. Therefore, development of CORC cable with a high strength core is highly desirable. Recently significant progress has been made in this area [17].

## IV. CONCLUSION

The critical current of a simple-stacked REBCO cable and a CORC® cable were measured using a SCT at 4.2 K in magnetic field up to 12 T. The resistance of each solder joint between SCT and simple-stacked cable is 3.7 nΩ. Critical current of the simple-stacked cable indicates irreversible damage by preceding measurements. The critical current of the CORC® cable was measured to be 23.7 kA at 5 T. Subsequent measurement showed sample quench indicative of electromechanical damage.

### ACKNOWLEDGMENT

The authors would like to thank Mr. Eric Arroyo for assistance in operation of the 12 T split magnet.

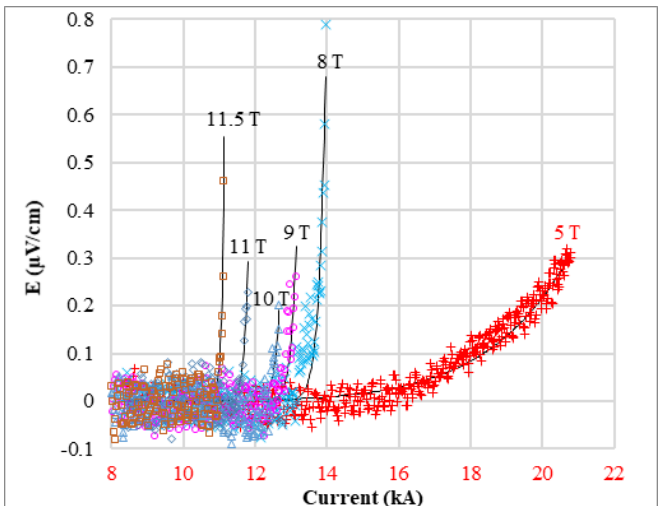

Fig. 7.  E-I curves of the CORC® sample at applied fields of 5 - 11.5 T.